\documentclass[aps,pre,floatfix,showpacs,twocolumn,groupedaddress]{revtex4}
\usepackage[english]{babel}
\usepackage{graphicx}
\usepackage{dcolumn}
\usepackage{amsmath}

\newcommand{\vs}{{\it vs.}}
\newcommand{\ie}{{i.e.}}

\begin{document}

\title{Structure and dynamics of hyaluronic acid semidilute solutions: a
dielectric spectroscopy study}

\author{T.\ Vuleti\'{c}}
\email{tvuletic@ifs.hr}
\homepage{http://real-science.ifs.hr/}

\affiliation{Institut za fiziku, Zagreb, Croatia}

\author{S.\ Dolanski Babi\'{c}}
\altaffiliation {Permanent address: Department of physics and biophysics,
Medical School, University of Zagreb, Croatia.}

\author{T.\ Ivek}
\author{D.\ Grgi\v{c}in}
\author{S.\ Tomi\'{c}}
\affiliation{Institut za fiziku, Zagreb, Croatia}

\author{R.\ Podgornik} \affiliation{Department of Physics, Faculty of
Mathematics and Physics and  Institute of Biophysics, School of Medicine,
University of Ljubljana,  Ljubljana, and J.\ Stefan Institute, Ljubljana,
Slovenia and \\ Laboratory of Physical and Structural Biology, Eunice Kennedy
Shriver National Institute of Child Health and Human Development, National
Institutes of Health, Bethesda MD, USA}

\date{\today}

\begin{abstract}
Dielectric spectroscopy is used to investigate fundamental length scales
describing the structure of hyaluronic acid sodium salt (Na-HA) semidilute
aqueous solutions. In salt-free regime, the length scale of the relaxation mode
detected in MHz range scales with HA concentration as $c_\mathrm{HA}^{-0.5}$ and
corresponds to the de Gennes-Pfeuty-Dobrynin correlation length of
polyelectrolytes in semidilute solution. The same scaling was observed for the
case of long, genomic DNA. Conversely, the length scale of the mode detected in
kHz range also varies with HA concentration as $c_\mathrm{HA}^{-0.5}$ which
differs from the case of DNA ($c_\mathrm{DNA}^{-0.25}$). The observed behavior
suggests that the relaxation in the kHz range reveals the de Gennes-Dobrynin
renormalized Debye screening length, and not the average size of the chain, as
the pertinent length scale. Similarly, with increasing added salt the
electrostatic contribution to the HA persistence length is observed to scale as
the Debye length, contrary to scaling pertinent to the Odijk-Skolnick-Fixman
electrostatic persistence length observed in the case of DNA. We argue that the
observed features of the kHz range relaxation are due to much weaker
electrostatic interactions that lead to the absence of Manning condensation as
well as a rather high flexibility of HA as compared to DNA.
\end{abstract}


\maketitle

\section{Introduction}

Dynamics of counterion atmosphere in the vicinity of polyelectrolytes can be used as an important structural tool in order to better understand their functional properties and is thus of fundamental importance also for various areas of bio-nanotechnology and bio-medical sciences in general. In biological context, charged
polyelectrolytes like deoxyribonucleic acid (DNA), ribonucleic acid (RNA),
polypeptides and polysaccharides such as hyaluronic acid (HA) are essential for
life and make their mark in many structural and functional aspects of the
cellular environment \cite{Daune03}. DNA is in many respects a paradigm of a
semiflexible highly charged polymer whose complex behavior was studied in great
detail. In aqueous solutions it assumes a conformation of an extended
statistical coil, whereas {\it in vivo} long genomic DNA is usually compactified
to fit within the micron-sized nucleus of eukaryotic cells or even smaller
nano-scale viral capsids \cite{Bloomfield00}.

HA is a member of the glycosaminoglycan family and is an alternating copolymer 
of D-glucuronic acid and N-acetyl glucosamine that occurs in connective tissue 
and mucous substances in the body. Similarly as in the case of DNA, HA always 
comes as a highly asymmetric salt with positive counterions, in our case as 
sodium salt Na-HA. In aqueous solutions its carboxyl groups are completely 
dissociated making HA a charged polyelectrolyte.  In the crystalline form, HA 
covers a broad range of conformations depending on pH, whereas in solution HA is
a single-stranded helix \cite{Elias}.

Parameters known to be relevant for the counterion dynamics in polyelectrolyte 
solutions include valency, strength of electrostatic interactions, concentration 
of polyions and added salt ions. Characteristic frequencies in collective 
counterion dynamics span several orders of magnitude ranging from kHz to MHz. 
These characteristic frequencies were found to be directly correlated with the 
details of the counterion motion around polyions 
\cite{Bordi,Ito,Tomic06,Tomic07,Tomic08}. For the sake of completeness we 
reiterate that in the polyelectrolyte solutions one also finds a relaxation 
related to water, confined to the GHz range  \cite{Bordi,Nandi}. For the 
strongly charged polyelectrolytes such as DNA one distinguishes  two distinct 
types of counterions: condensed counterions which are tightly bound to the 
polyions and free counterions. These 
counterion types  are considered transient, meaning that there is a constant 
dynamic exchange between them. Relaxation mode arising from the collective 
motion of condensed counterions takes place in the kHz frequency range (low
frequency, LF mode), whereas the mode associated with free counterion motion is
detected in the higher MHz frequency range (high frequency, HF mode). In the
linear regime, \ie{}, for small applied ac electric fields, these relaxation
modes probe directly the conformational features of either a single polyelectrolyte
chain or the structure of an ensemble of many chains in solution. Our dielectric
spectroscopy (DS) measurements on long DNA semidilute solutions \cite{Tomic06,Tomic07}
show that the measured fundamental length scale, probed by the LF mode, is equal
to the size of the Gaussian chain composed of correlation blobs that scales as 
$c_\mathrm{DNA}^{-0.25}$ in the low added salt limit. In the high added salt 
limit the LF mode characteristic length scale goes over to the
Odijk-Skolnick-Fixman (OSF) electrostatic persistence length $L_\mathrm{p}$ and
scales as $L_\mathrm{p} = L_0 + a I_\mathrm{s}^{-1}$. Here $L_0$ is the DNA
structural persistence length of 50~nm and $I_\mathrm{s}$ is the ionic strength
of the added salt. On the other hand, the HF mode is probing the collective
properties of the DNA solution which are characterized by the
de Gennes-Pfeuty-Dobrynin (dGPD) correlation length or the solution mesh size,
that scales as $c_\mathrm{DNA}^{-0.5}$. This result was equally reported on
diverse synthetic polyelectrolytes \cite{Bordi,Ito}.

An intriguing issue is if and how the dielectric relaxation modes will change 
for polyelectrolytes, such as HA, which are much more flexible than DNA,
\ie{}, how is the counterion dynamics affected  when the structural
persistence length becomes smaller than the dominating correlation length. We 
note that the double-stranded DNA (ds-DNA) is best described as a strongly 
charged semi- flexible polyelectrolyte whose structural persistence length, 
$L_0$ close to 50~nm, is larger than the correlation length. Another interesting 
issue is what happens in the case of weakly charged polyelectrolytes, again such 
as HA, whose Manning charge density parameter $\eta < 1$, so that according to 
the Manning criterion no counterion condensation takes place. Here $\eta = z 
l_\mathrm{B} / b$, where $z$ is the valency of the counterion, $b$ is the linear 
charge spacing and $l_\mathrm{B}$ is the Bjerrum length. The question then 
arises whether or not free counterions, in addition to their primary role, take 
over the role of condensed counterions as in strongly charged polyelectrolytes. 
In other words, do free counterions, in addition to oscillating between the 
correlated polyion chains in the solution, also perform an oscillatory motion 
along the polyion chain revealing in this way the conformational features of a 
single chain. Previous DNA studies have shown that the LF relaxation mode does 
not allow a clearcut separation between condensed and free counterions as 
relaxation entities and that both of them contribute to various extent in 
different salt and DNA concentration regimes \cite{Tomic07}.

In this work we address these issues by characterizing the dynamic behavior of 
semidilute solutions of a sodium salt of hyaluronic acid (Na-HA). HA is a weakly 
charged polyelectrolyte whose disaccharide monomer is 1~nm long implying the 
Manning charge density parameter $\eta \approx 0.71$. The structural persistence 
length is about 9~nm \cite{Buhler04} and is larger than the monomer size so 
that, according to the standard criterion, HA belongs to semiflexible 
polyelectrolytes, being much stiffer than the standard synthetic polyelectrolyte 
polystyrene sulfonate Na-PSS, but at the same time much more flexible than
ds-DNA.

\section{Materials and Methods}

In this study we used hyaluronic acid sodium salt from streptococcus equi
sp.\ Fluka 53747 obtained from Sigma-Aldrich. The low protein content $\leq 1$ 
is declared by the manufacturer. An average molecular weight is about
$1.63 \times 10^6$~Daltons, implying 4000 monomers in average. For HA, 
disaccharide monomer is of molecular weight 401~g/mol and there is one Na$^+$ 
ion per HA monomer. This means that the molar concentrations of HA monomers, as 
well as Na$^+$ counterions are related by
$c_\mathrm{in} \left[\textrm{mM}\right] = c_\mathrm{HA} \left[\textrm{mg/mL}\right] \times 2.5$~$\mu$mol/mg
to the HA concentration by weight. We perform a systematic study of how the 
dielectric properties of these polydisperse long HA fragments in aqueous 
solutions evolve upon change of HA concentration and added salt over a range of 
two orders of magnitude. HA solutions were prepared as described previously (see 
Materials and Methods I in Ref.\ \cite{Tomic07}, preparation protocols I and 
II.1). A crude estimate of the crossover concentration between dilute and 
semidilute regime $c^\ast$ based on the de Gennes arguments \cite{deGennes76}, 
and taking into account that the average HA fragments are 4~$\mu$m long, yields 
$c^\ast$ of the order of $0.00004$~mg/mL, which is more than two orders of 
magnitude below the lower concentration bound in our experiments \cite{note1}. 
This means that we are effectively always in semidilute regime which has been 
studied in depth experimentally in the case of long  polydisperse DNA fragments 
in aqueous solutions \cite{Tomic07}.

\subsection{Capacitance chamber and the electrode polarization}

Dielectric spectroscopy up to 100 MHz is usually performed with a capacitance 
chamber \cite{Schwan,Essex,Agilent,Holling,Roldan}. There, a sample solution is 
applied between the electrodes of a parallel plate capacitor and a frequency 
dependent impedance is measured. We performed our measurements with an in-house 
designed chamber, with built-in temperature control. The chamber was connected 
in a 4-terminal pair configuration to an Agilent 4294A precision impedance 
analyzer \cite{Agilent}. The measurement functions were conductance $G(\omega)$ 
and the capacitance in parallel $C_p(\omega)$, where $\omega=2\pi\nu$ is frequency. 
These combine into complex admittance (inverse of complex impedance $Z$):
\begin{equation}
Y(\omega)=G(\omega)+\imath \omega C_p(\omega)  
\label{YGCp}
\end{equation}
Admittance is sampled at 201 frequencies, \ie{}, at 33 points per frequency 
decade for the frequency range 100Hz-110 MHz. A single frequency point is 
sampled with bandwidth BW=5 setting (longer acquisition time, but lower noise -
BW=1 would be faster, however more noisy, and would demand averaging 10 times for one 
point). In addition, three consecutive frequency sweeps are taken in order to 
average out the temperature variations. Total time for the spectrum measurement 
amounts to 120~s.

In the measurements with a capacitance chamber a well-known electrode 
polarization (EP) influences the results. EP is due to the rearrangement of ions 
and a build-up of spatial charge in the vicinity of the electrodes, which 
manifests itself as a large frequency dependent capacitance, $C_{EP}$, in series 
with the sample chamber impedance -  impedance of a capacitor containing a 
conducting dielectric. EP can be roughly represented in kHz range (for simple 
electrolytes at concentrations below 10  mM) by
$C_{EP} = a \cdot \omega^{\alpha}$, where  $\alpha \approx -2$. EP also strongly
reduces measured conductance, $G$ in the same low-frequency range. The effect is
proportional to the ionic conductivity of the electrolyte, \ie{}, to its ionic
strength $I_s$. The contributions (of polarization currents) due to the system
of interest (e.g., polyelectrolyte, colloid) may be extracted only in the
frequency range where they are of comparable or larger magnitude than EP effects.
This practically limits dielectric spectroscopy below 100 MHz to studies of
samples in electrolytes with  $I_s$ not above 10 mM \cite{Schwan,Essex,Holling,Roldan}.

Various experimental methods \cite{Bordi} aim to reduce the EP effect, e.g.,
4-contact measurement cells \cite{Schwan68}, electromagnetic induction measurement 
methods or simple utilization of the microporous, platinum black electrodes 
(however, Pt-black deteriorates with repeated usage \cite{Holling}). A different 
approach is to model the EP contribution to the sample impedance and 
consequently remove it. However, while this may be done in theory, in practice, 
every new polyelectrolyte solution presents itself as a system where the EP 
effect can have a different influence, and our incomplete understanding of this 
phenomenon reappears \cite{Roldan}. Even for a simple capacitance chamber there 
remain two experimental approaches for reduction of EP. The first one is to make 
measurements at different electrode separations \cite{Holling}. Then, as EP effect is the only one dependent on the electrode surface area as opposed to the volume- dependent sample contribution, one can distinguish the two contributions. 
The second EP reduction approach is the reference subtraction method 
\cite{Bordi,Davey, Saif91,Grosse} which is used in our work. In addition to HA samples, the reference 
samples of a simple electrolyte NaCl were also measured and corresponding 
spectra subtracted from HA spectra in order to minimize stray impedances, 
including the free ion contribution and EP effects, and extract the response due 
to HA only \cite{Tomic07}. Naturally, this method has limitations in the sense 
that the EP effects are not necessarily the same for the sample and reference 
solutions - an issue similar to the abovementioned EP modeling methods \cite{Bordi}.

While recognizing its limitations, we have chosen the reference subtraction
method, and designed our setup accordingly, as it allows for quick data analysis
and high sample throughput (a sample is exchanged every 10 minutes) with a
minimal sample volume (100~$\mu$L).

\begin{figure}
\includegraphics[clip,width=1.0\linewidth]{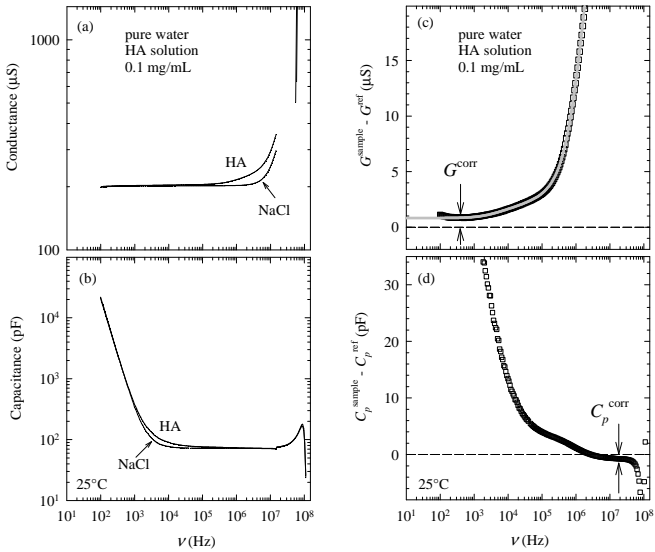}
\caption{\label{Fig1} (a),(b) Double logarithmic plot of the frequency dependence
of the (a) conductance  and capacitance  for 0.1~mg/mL
HA solution and the matching reference 0.14 mM NaCl solution.
(c),(d) Frequency dependence of the {\em differential} admittance components
$G(\omega)=G^\mathrm{sample}(\omega)-G^\mathrm{ref}(\omega)$ and
$C_p(\omega)=C_p^\mathrm{sample}(\omega)-C_p^\mathrm{ref}(\omega)$.
The correction constants are denoted $G^\mathrm{corr}$ and $C_p^\mathrm{corr}$.
The full line in panel (c) is the $G(\omega)$ spectrum calculated from the complex
dielectric function fit (see text). }
\end{figure}

Fig.\ \ref{Fig1}(a) and (b) shows the 0.1~mg/mL HA pure water solution 
conductance and capacitance spectra in conjunction with data for 0.14~mM NaCl 
reference solution of matching conductance (matched at 100~kHz). The 
corresponding curves show only a small mismatch which indicates that the main 
contributions come from the free ion conduction or the EP effect, while the 
contribution of interest $G(\omega)$, $C_p(\omega)$, which is due to the polyion 
and its counterion atmosphere, is minor. Therefore, assuming that the conductivity contribution of each entity in the solution is additive \cite{Osawa71}, we have treated the contribution of interest additively:  
\begin{eqnarray}
G(\omega)=G^\mathrm{sample}(\omega)-G^\mathrm{ref}(\omega)\\
C_p(\omega)=C_p^\mathrm{sample}(\omega)-C_p^\mathrm{ref}(\omega)
\label{GCp-GCp}
\end{eqnarray}
and Fig.\ \ref{Fig1}(c),(d) shows the result of this subtraction. The 
contribution of the system of interest (\ie{}, the polyion and its counterion 
atmosphere) to conductivity $\sigma(\omega)=K \cdot [G(\omega)+\imath \omega C_p(\omega)]$
can be used to express its contribution to the complex dielectric function
$\varepsilon(\omega) = \varepsilon^{\prime}(\omega) - \imath \varepsilon^{\prime\prime}(\omega) = \sigma(\omega)/( \imath \omega \varepsilon_0)$
(here $K$ is the geometrical sample chamber constant $K=l/S$, $l$ is the
electrode separation and $S$ the surface area, and $\varepsilon_0$ is the
permittivity of vacuum). We arrive to the final expression for the imaginary and 
real part of dielectric function:
\begin{eqnarray}
\varepsilon^{\prime\prime}(\omega)=\frac{K\cdot\left(G^\mathrm{sample}(\omega)-G^\mathrm{ref}(\omega)-G^\mathrm{corr}\right)}{\omega\varepsilon_0} \label{epseps-G}
\\
\varepsilon^{\prime}(\omega)=\frac{K\cdot\left(C_p^\mathrm{sample}(\omega)-C_p^\mathrm{ref}(\omega)-C_p^\mathrm{corr}\right)}{\varepsilon_0}
\label{epseps-Cp}
\end{eqnarray}
Due to the imperfect matching of the reference solution small corrections
$G^\mathrm{corr}$ and $C_p^\mathrm{corr}$ remain. They are read from the data in Fig.\ \ref{Fig1} (c) and (d) and taken into account in the subsequent fitting procedure.

%%%%%%%%%%%%%%FIGURE%%%%%%%%%%%%%%%%%%%%%%%%
%%%%%%%%%%%%%%FIGURE%%%%%%%%%%%%%%%%%%%%%%%%

\subsection{Complex dielectric function - fit to Cole-Cole expression}

\begin{figure}
\includegraphics[clip,width=0.6\linewidth]{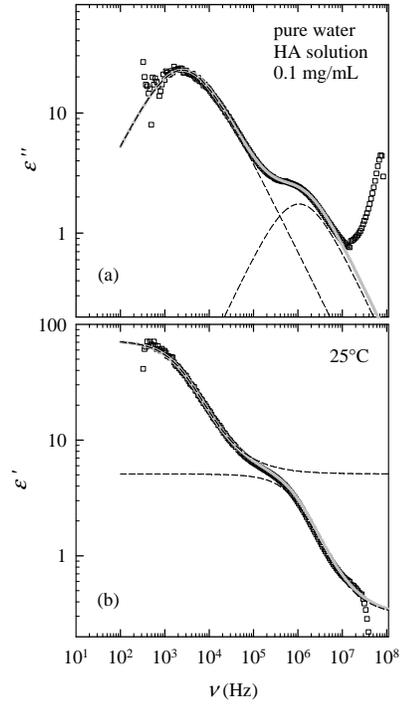}
\caption{\label{Fig2ab}
Double logarithmic plot of the frequency dependence of the
 imaginary ($\varepsilon^{\prime\prime}$) and real ($\varepsilon^\prime$) part
 of the dielectric function for the 0.1~mg/mL HA pure water solution. The full
 lines are fits to the sum of the two Cole-Cole forms (see text); the dashed
 lines represent a single form.}
\end{figure}

In Fig.\ \ref{Fig2ab} we show the frequency-dependent imaginary and real 
part of the dielectric function for the 0.1 mg/mL HA pure water solution. These 
spectra are calculated according to Eqs.\ref{epseps-G} and \ref{epseps-Cp} from the $G$ and $C_p$ 
spectra [Fig.\ \ref{Fig1}(c),(d)]. The resulting spectra cover the frequency 
window from 500 Hz to 30 MHz. The reference subtraction procedure has reliably 
removed the influence of electrode polarization, visible in the raw 
$C_p^\mathrm{sample}$ spectra up to 30 kHz, down to 500 Hz. The stray impedance 
effects noticeable above 10 MHz have shown to be more resilient - the 
subtraction procedure removed them only up to 30 MHz. We note that the 
capability of the reference subtraction procedure to remove the EP effects 
diminishes (low frequency bound increases) with increasing concentration and 
total sample conductivity. For this reason the highest successfully analyzed
Na-HA concentration was 1.25 mg/mL.

\begin{figure}
\includegraphics[clip,width=0.6\linewidth]{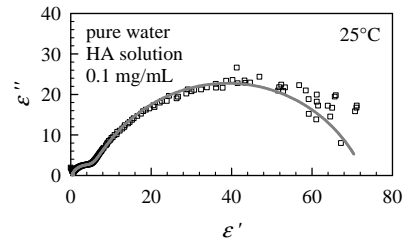}
\caption{\label{Fig2c} Cole-Cole plot of the dielectric response for the
0.1~mg/mL HA pure water solution. The full line is a fit to the sum of the two
Cole-Cole forms (see text). The LF mode contributes as the dominant arch, while
the smaller HF mode is found near the origin of the axes.}
\end{figure}

The Cole-Cole function has been widely and successfully used to describe 
relaxation processes in disordered systems. The dielectric function spectra 
consist of two broad modes and the data can only be successfully fitted to a sum 
of two Cole-Cole forms
\begin{equation}
\varepsilon(\omega)-\varepsilon_\infty
 = \frac{\Delta\varepsilon_{\mathrm{LF}}}{[ 1 + \left(i \omega \tau_{0,\mathrm{LF}} \right)^{ 1-\alpha_{\mathrm{LF}} } ]}
 + \frac{\Delta\varepsilon_{\mathrm{HF}}}{[ 1 + \left(i \omega \tau_{0,\mathrm{HF}} \right)^{ 1-\alpha_{\mathrm{HF}} } ]}
\label{Cole}
\end{equation}
where $\varepsilon_\infty$ is the high-frequency dielectric constant, 
$\Delta\varepsilon$ is the dielectric strength, $\tau_0$ the mean relaxation 
time and $1-\alpha$ the symmetric broadening of the relaxation time distribution 
function of the LF and HF dielectric mode. 
Measured data were analyzed by using the least-squares method in the complex 
plane: the same set of parameters fits both the real and imaginary spectra. 
Besides improving our ability to resolve two modes, this also provides an 
important Kramers-Kronig consistency check for the experimental data obtained 
indirectly by the reference subtraction procedure. This consistency is 
demonstrated with a Cole-Cole plot (Fig.\ \ref{Fig2c}). The larger arch in this 
figure corresponds to the LF mode, which is more pronounced than the HF one 
found near the origin of the axes. The final verification of our fits is 
comparing the $G(\omega)$ data to the conductance spectrum calculated from the 
dielectric fit parameters using the inverse of the expression \ref{epseps-G} 
[Fig.\ \ref{Fig1}(c), full line].

%%%%%%%%%%%%%%FIGURE%%%%%%%%%%%%%%%%%%%%%%%%
%%%%%%%%%%%%%%FIGURE%%%%%%%%%%%%%%%%%%%%%%%%

%%%%%%%%%%%%%%FIGURE%%%%%%%%%%%%%%%%%%%%%%%%
%%%%%%%%%%%%%%FIGURE%%%%%%%%%%%%%%%%%%%%%%%%

\section{Results}

\begin{figure}
\includegraphics[clip,width=1.00\linewidth]{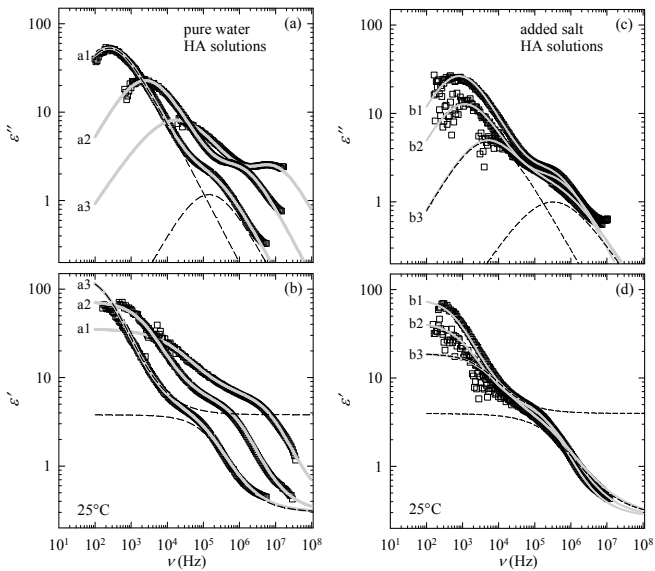}
\caption{\label{Fig3}Double logarithmic plot of the frequency dependence of the
imaginary ($\varepsilon''$) and real ($\varepsilon'$) part of the dielectric
function at $T=25^\circ$C of (a), (b) pure water HA solutions for representative
a1-a3 (1, 0.1, 0.0125~mg/mL) HA concentrations and
(c), (d) HA solutions of concentration $c_\mathrm{HA}=0.03$~mg/mL for
three representative b1-b3 (0, 0.12, 0.25~mM) added salt concentrations. The
full lines are fits to the sum of the two Cole-Cole forms (see text);
the dashed lines represent a single form.}
\end{figure}

Fig.\ \ref{Fig3} shows the frequency-dependent imaginary and real part of the
dielectric function for HA aqueous solutions with selected HA concentrations.
The results for pure water HA solutions (HA concentrations $\mathrm{a1} = 1$~mg/mL,
$\mathrm{a2} = 0.1$~mg/mL, $\mathrm{a3} = 0.0125$~mg/mL) are shown in panels (a)
and (b), while the results for solutions of 0.03~mg/mL HA concentration with several
different added salt concentrations
($\mathrm{b1} = 0$~mM, $\mathrm{b2} = 0.12$~mM, $\mathrm{b3} = 0.25$~mM)
are shown in panels (c) and (d). The observed dielectric response is complex and 
comparable to the one previously observed in DNA solutions. The two broad modes 
show a symmetrical broadening of the relaxation time distribution function 
described by the parameter $1-\alpha \approx 0.8$. The mode centered at higher 
frequencies, 0.15~MHz $ < \nu_\mathrm{HF} < $ 15~MHz, is characterized by 
dielectric strength $3  < \Delta\varepsilon_\mathrm{HF} < 5.5 $. The other mode 
is larger, $20 < \Delta\varepsilon_\mathrm{LF} < 150$, and centered at lower 
frequencies, 0.3~kHz $< \nu_\mathrm{LF} <$ 50~kHz.

The polarization response of polyelectrolyte solutions in the kHz-MHz range is
due to oscillations of counterions induced by an applied ac field. Since the
counterion displacement happens by diffusion, the dielectric response is
basically characterized by the mean relaxation time
\begin{equation}
\tau_0 \propto L^2/D_\mathrm{in}
\label{tauLD}
\end{equation}
Here $L$ is the associated characteristic
length scale, and $D_\mathrm{in}$ is the diffusion constant of counterions \cite{OBrian86}.
Experimental data \cite{Wong} and theoretical estimates \cite{Bordi02} show that
the renormalization of the diffusion constant of bulk ions due to the presence of
polyions is negligible, leading to a value of
$D_\mathrm{in}=1.33 \times 10^{-9}$~m$^2$/s for Na$^+$ counterions. In other
words, the Cole-Cole fits allow us to extract the characteristic time $\tau_0$
and calculate the corresponding length scale for each of the relaxation modes.

%%%%%%%%%%%%%%FIGURE%%%%%%%%%%%%%%%%%%%%%%%%
%%%%%%%%%%%%%%FIGURE%%%%%%%%%%%%%%%%%%%%%%%%

While the length scale describes the counterion displacement, the dielectric 
strength of the relaxation modes is related to the polarizability due to the
counterion displacement. The polarizability $\alpha$ is estimated 
to be proportional to the square of the displacement \cite{Ito,Bordi}, \ie{}, to 
the square of the characteristic length scale of the relaxation:
\begin{equation}
\alpha \propto l_B \cdot L^2
\label{alpha}
\end{equation}
This scaling form stems from the linear response arguments already invoked by Odijk 
\cite{Odijk79}. Taking into account the number of the relaxing entities, a 
qualitative expression is reached for the dielectric strength due to the 
polarizability of the counterion atmosphere around a polyion in 
polyelectrolytes:
\begin{equation}
\Delta\varepsilon \propto f \cdot c \cdot \alpha
\label{increment}
\end{equation}
Combining Eqs.\  \ref{alpha} and  \ref{increment} we get
\begin{equation}
\Delta\varepsilon \propto f \cdot c \cdot  l_B \cdot L^2
\label{increment2}
\end{equation}
Here $f$ denotes the fraction of the total counterions of concentration c that take part in a given 
relaxation. It is important to understand here that the dielectric strength of a 
mode is primarily defined by the length scale revealed by the mode. The 
concentration dependence of the fraction of counterions may be deduced by the 
ratio 
\begin{equation}
f \propto \Delta\varepsilon /(c \cdot L^2)
\label{ratio_f}
\end{equation}
which is solely based on the experimentally obtained parameters of the dielectric modes.

Similarly complex spectra as shown in Fig.\ \ref{Fig3} have also been observed for long polydisperse DNA 
semidilute solutions \cite{Tomic06,Tomic07}. However, the monomer concentration 
and added salt dependence measured for HA are rather distinct, indicating that 
mechanisms of counterion relaxation for HA as compared to long DNA solutions are 
not identical. Two origins of these differences might be envisaged. First is the 
rather weak fixed charge on the HA molecule leading to the absence of Manning 
condensation and second is a much higher flexibility of the HA, whose 
persistence length is about 9~nm as opposed to 50~nm for the ds-DNA.

\subsection{HF mode}

\begin{figure}
\includegraphics[clip,width=0.8\linewidth]{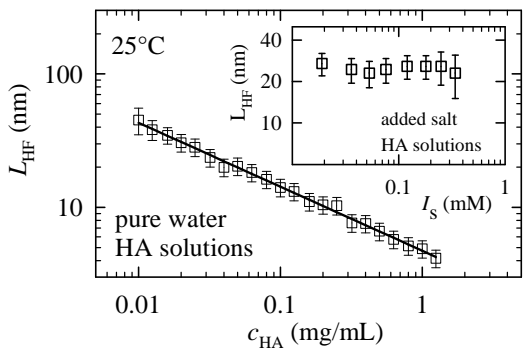}
\caption{\label{Fig4}
Main panel: Characteristic length of the HF mode ($L_\mathrm{HF}$) for pure
water HA solutions as a function of HA concentration ($c_\mathrm{HA}$). The full
line is a fit to the power law $L_\mathrm{HF} \propto
c_\mathrm{HA}^{-0.48\pm0.02}$. Inset:  $L_\mathrm{HF}$ for HA solutions with
varying added salt ($I_\mathrm{s}$) for a representative HA concentration
$c_\mathrm{HA}=0.03$~mg/mL.
}
\end{figure}

We first describe the characteristics of the HF mode. For pure water HA 
solutions the characteristic length $L_\mathrm{HF}$ decreases with increasing HA 
concentration in two decades wide concentration range following the power law 
$L_\mathrm{HF} \propto c_\mathrm{HA}^{-0.48\pm0.02}$ as a function of the HA 
concentration (main panel of Fig.\ \ref{Fig4}). This dependence $-0.48\pm0.02$ 
suggests that in this regime $L_\mathrm{HF}$ is proportional to the solution 
mesh size or the de Gennes-Pfeuty-Dobrynin (dGPD) correlation length $\xi$ that 
scales as $c_\mathrm{HA}^{-0.5}$, as theoretically expected for the salt-free 
semidilute solutions of flexible polyelectrolytes 
\cite{Bordi,deGennes76,Pfeuty78,Dobrynin}:
\begin{equation}
\xi \propto f ^{-2/7}\cdot c ^{-1/2}
\label{xif2c}
\end{equation}
We note that the fraction of the counterions $f$, taking part in the relaxation, is here defined by the Manning charge density parameter $\eta$ as  $f=1/\eta=b/l_B$ and is concentration-independent. Therefore, Eq.\ \ref{xif2c} reduces to $\xi \propto c ^{-1/2}$.
Finally, we note that the same fundamental length  scale is also pertinent for semiflexible chains like DNA  \cite{Tomic06,Tomic07}, as previously predicted by Odijk \cite{Odijk79}.

%%%%%%%%%%%%%%FIGURE%%%%%%%%%%%%%%%%%%%%%%%%
%%%%%%%%%%%%%%FIGURE%%%%%%%%%%%%%%%%%%%%%%%%

A set of data (inset of Fig.\ \ref{Fig4}) for $c_\mathrm{HA}=0.03$~mg/mL with
varying added salt concentrations shows that the de Gennes-Pfeuty-Dobrynin
behavior of $L_\mathrm{HF}$ remains unchanged at least for all $I_\mathrm{s} <
0.34$~mM. Unfortunately for $I_\mathrm{s} > 0.34$~mM the accuracy of the data
becomes rather unreliable due to progressive merging of the HF and LF modes at
these added salt levels making it thus impossible to distinguish the two modes
with high enough accuracy. This result nevertheless shows that the dGPD
correlation length remains relevant length scale of the HF mode which indeed
depends only on the polymer concentration and not on the added salt. It is
noteworthy that this holds up to rather high added salt levels as compared to HA
counterion concentrations, that is at least up to $2I_\mathrm{s} \approx 9
c_\mathrm{in}$.

\begin{figure}
\includegraphics[clip,width=0.8\linewidth]{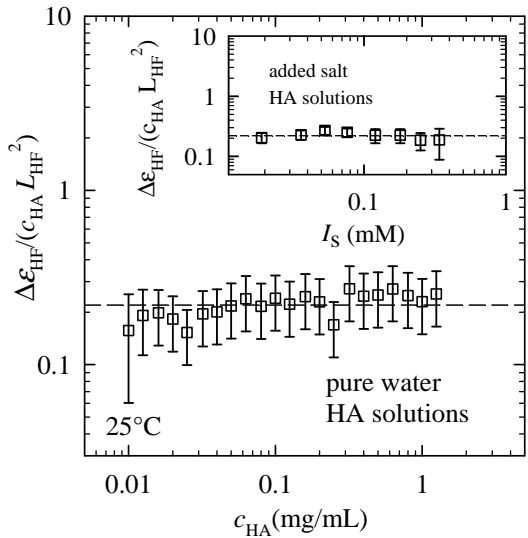}
\caption{\label{Fig5}
Main panel: Normalized dielectric strength of the HF mode 
$\Delta\varepsilon_\mathrm{HF} /(c_\mathrm{HA} \cdot L_\mathrm{HF}^2)$ as a 
function of HA concentration $c_\mathrm{HA}$ for pure water HA solutions. 
Inset:  $\Delta\varepsilon_\mathrm{HF} /(c_\mathrm{HA} \cdot L_\mathrm{HF}^2)$ 
for HA solutions with varying added salt ($I_\mathrm{s}$) for a representative 
HA concentration $c_\mathrm{HA}=0.03$~mg/mL. The dashed lines are guides for 
the eye.}
\end{figure}

The fraction of counterions $f_\mathrm{HF}$ participating in the HF process is 
proportional to the normalized dielectric strength 
$\Delta\varepsilon_\mathrm{HF} /(c_\mathrm{HA} \cdot L_\mathrm{HF}^2)$ (see          Eq.\ \ref{ratio_f}). In the 
main panel of Fig.\ \ref{Fig5} we show the dependence of 
$\Delta\varepsilon_\mathrm{HF} /(c_\mathrm{HA} \cdot L_\mathrm{HF}^2)$ on HA 
concentration in pure water solutions. In the inset we show the dependence on 
the ionic strength of added salt ions for a representative HA concentration 
$c_\mathrm{HA}=0.03$~mg/mL. The data indicate that the 
fraction of counterions participating in the HF relaxation process does not 
depend on the concentration of either HA or added salt in the concentration 
range studied. Here we point out a similarity with the semidilute DNA solutions 
where the relaxation mode related to the dGPD correlation length occurs in the 
same concentration and frequency ranges and features qualitatively similar, 
concentration-independent $f_\mathrm{HF}$ \cite{Tomic07}. 
This result validates the standard theoretical models which use the Manning-based definition of $f$ as the concentration-independent parameter.  Next, it is worth noting that for DNA  
$f_{\textrm{HF}}$  remains constant when salt is added to the solution
only as long as polyion concentration is substantially larger than 
$I_s$.  However, as soon as the concentration of added salt ions prevails over 
the concentration of intrinsic counterions, $f_{\textrm{HF}}$ starts to 
decrease. Whereas this effect was clearly observed for DNA solutions, in the case of HA it can only be guessed due to smaller accuracy at added salt concentrations larger than 0.2~mM \cite{note2}.

%%%%%%%%%%%%%%FIGURE%%%%%%%%%%%%%%%%%%%%%%%%
%%%%%%%%%%%%%%FIGURE%%%%%%%%%%%%%%%%%%%%%%%%

\subsection{LF mode}

\begin{figure}
\includegraphics[clip,width=0.8\linewidth]{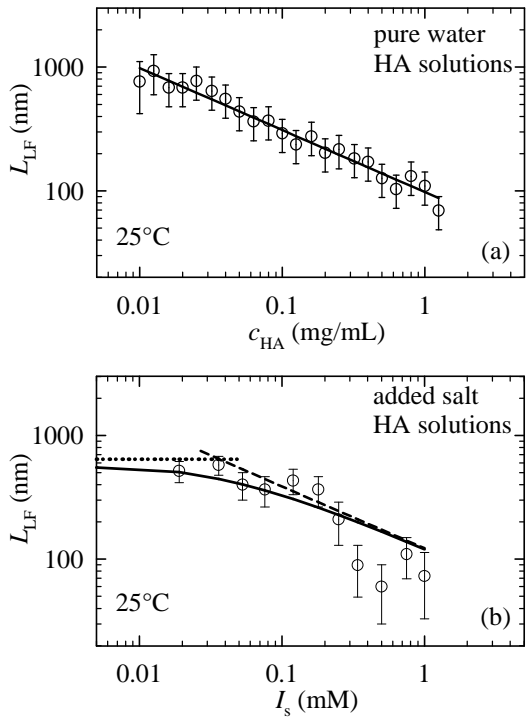}
\caption{\label{Fig6}(a) Characteristic length of the LF mode ($L_\mathrm{LF}$)
for pure water HA solutions as a function of HA concentration ($c_\mathrm{HA}$).
The full line is a fit to the power law
$r_\mathrm{B} = C \cdot (B/bc_\mathrm{HA})^{0.5\pm0.02}$ with $C = 4.3$ (see
text).
(b) $L_\mathrm{LF}$ for HA solutions with
varying added salt ($I_\mathrm{s}$) for a representative HA concentration:
$c_\mathrm{HA}=0.03$~mg/mL. The full line is a fit to the expression
$r_\mathrm{scr} = C \cdot \left[B /
\left(b(c_\mathrm{HA}+2AI_\mathrm{s})\right)\right]^{0.5}$
with $C = 4.3$ (see text). The dashed line corresponds to
$L_\mathrm{p} \propto I_\mathrm{s}^{0.5}$. The dotted line denotes the value of
$L_\mathrm{LF}$ in salt-free limit for $c_\mathrm{HA}=0.03$~mg/mL.}
\end{figure}

We now address the LF mode. For pure water HA solutions, the characteristic
length $L_\mathrm{LF}$ decreases with increasing HA concentration in two decades
wide concentration range [Fig.\ \ref{Fig6}(a)] following the power law
$L_\mathrm{LF} \propto c_\mathrm{HA}^{-0.5\pm0.02}$. This scaling behavior
differs from the one observed for long DNA \cite{Tomic06,Tomic07}, where
$L_\mathrm{LF}$ was identified with the average size of the Gaussian chain that
behaves as a random walk of correlation blobs and scales as $R \propto
c_\mathrm{DNA}^{-0.25 }$ \cite{deGennes76,Dobrynin}. In the case of HA solutions
the observed dependence of $L_\mathrm{LF}$ can be fit nicely to the de
Gennes-Dobrynin (dGD) renormalized Debye screening length which scales as
\begin{equation}
r_\mathrm{B} = C \cdot (B / b c_\mathrm{HA})^{0.5}
\label{r_B}
\end{equation}
 in the salt-free regime
\cite{Dobrynin}, where we get $C = 4.3$. Here the screening is due to HA counterions and the renormalization takes into
account the polyion chain properties. $r_\mathrm{B}$ has the same scaling as the Debye length, but is larger in
magnitude. $B$ is a parameter which is defined
as the ratio of the contour length $L_\mathrm{c}$ and the actual size $L$, being close to one for HA, while the monomer size $b$ is
1~nm. Importantly, $B$ is also defined {\em via} Manning concentration-independent parameter ($\eta=1/f$) and reads $B=(A^2 \cdot f)^{2/7}$. $A$ is the average number of monomers between charges. For HA there is no Manning condensation and $A = 1$. Now we can write
\begin{equation}
r_B = C \cdot f ^{1/7}\cdot (b c_\textrm{HA}) ^{-1/2}
\label{r_Bfc}
\end{equation}
which reduces to $r_B \propto c_\textrm{HA} ^{-1/2}$.
Dobrynin {\it et al.}\ \cite{Dobrynin} give only a lower bound of
$r_\mathrm{B}$, so that the numerical factor $C$ remains unknown in their
calculation. We note that in our previous dielectric spectroscopy experiments
\cite{Tomic06,Tomic07,Tomic08} various length scales correspond to the
theoretically expected values not only qualitatively but also
quantitatively, \ie{}, without any rescaling and prefactors \cite{note3}.
Therefore we take the numerical factor $C = 4.3$ as the upper bound up to which
the electrostatic screening length expands.

%%%%%%%%%%%%%%FIGURE%%%%%%%%%%%%%%%%%%%%%%%%
%%%%%%%%%%%%%%FIGURE%%%%%%%%%%%%%%%%%%%%%%%%

The dependence of $L_\mathrm{LF}$ on the added salt ionic strength
$I_\mathrm{s}$ is shown in Fig.\ \ref{Fig6}(b) for $c_\mathrm{HA}=0.03$~mg/mL.
The observed data can be nicely fit to the expression for the electrostatic
screening length  of the form
\begin{equation}
r_\mathrm{scr} = C \cdot \left[B /
\left(b(c_\mathrm{HA}+2AI_\mathrm{s})\right)\right]^{0.5}
\label{r_BIs}
\end{equation}
with $C = 4.3$. Here the electrostatic screening length is almost proportional
to the Debye length as first assumed by de Gennes
{\it et al.}\ \cite{deGennes76} and Dobrynin {\it et al.}\ \cite{Dobrynin} for
flexible polyelectrolytes. As before, $B = 1$ and $A = 1$. We believe that this fit reliably describes the
experimental data in spite of poorer accuracy of the data for added salt
concentrations close to 1~mM \cite{note2}. It is noteworthy that the observed
behavior of $L_\mathrm{LF}$,  that is of the deGennes-Dobrynin (dGD)
electrostatic screening length $r_\mathrm{scr}$, can be decoupled into high and
low salt regimes. In the former, $L_\mathrm{LF}$ is dominated by the influence
of the added salt ions [as shown by dashed line in Fig.\ \ref{Fig6}(b)], whereas
the low salt data level off with the limiting value corresponding to the value
of $L_\mathrm{LF}$ found in the salt-free solutions for this HA concentration
[see Fig.\ \ref{Fig6}(a)]. This observation indicates how added salt ions and HA
counterions compete for the dominant role in the screening of HA polyions
revealing at the same time the competition between two length scales describing
screening due to two different origins, just as in the case of DNA
\cite{Tomic06,Tomic07}.
At vanishing salt $L_\mathrm{LF}$ can be identified as the renormalized Debye
screening length due to HA counterions only
$r_\mathrm{B} = C \cdot (B/bc_\mathrm{HA})^{0.5}$, while, by analogy with DNA,
for finite added salt $L_\mathrm{LF}$  can be identified as the electrostatic
persistence length that apparently depends linearly on the Debye length and can
be fitted to the form $L_\mathrm{p} \propto I_\mathrm{s}^{-0.5}$. It is
noteworthy that the power law which thus describes the electrostatic persistence
length for HA differs from the one found for long DNA chains
\cite{Tomic06,Tomic07},
which correponds to the Odijk-Skolnick-Fixman (OSF) electrostatic persistence
length scaling as $L_\mathrm{p} \propto I_\mathrm{s}^{-1}$
\cite{Odijk77,Skolnick77}.

%%%%%%%%%%%%%%FIGURE%%%%%%%%%%%%%%%%%%%%%%%%
\begin{figure}
\includegraphics[clip,width=0.8\linewidth]{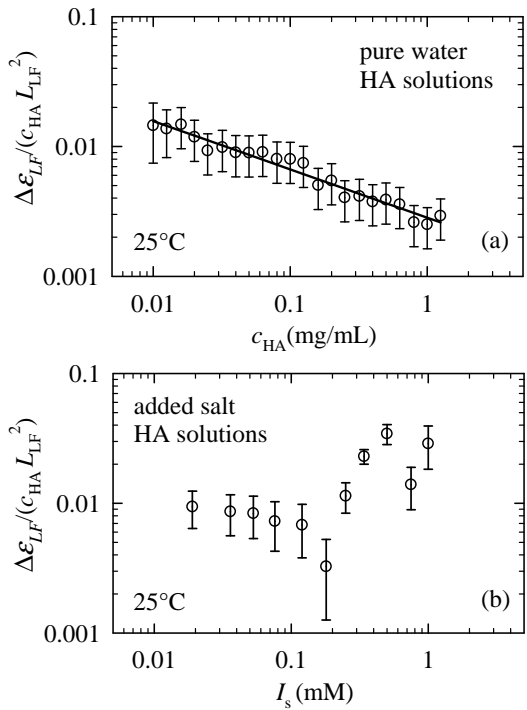}
\caption{\label{Fig7}
(a) Normalized dielectric strength of the LF mode
$\Delta\varepsilon_\mathrm{LF} /(c_\mathrm{HA} \cdot L_\mathrm{LF}^2)$ as a function of HA concentration $c_\mathrm{HA}$ for
pure water HA solutions. The full
line is a fit to the power law $\Delta\varepsilon_\mathrm{LF} /(c_\mathrm{HA} \cdot L_\mathrm{LF}^2) \propto
c_\mathrm{HA}^{-0.37\pm0.04}$. (b)  $\Delta\varepsilon_\mathrm{LF} /(c_\mathrm{HA} \cdot L_\mathrm{LF}^2)$ for HA solutions with
varying added salt ($I_\mathrm{s}$) for a representative HA concentration
$c_\mathrm{HA}=0.03$~mg/mL.
}
\end{figure}
%%%%%%%%%%%%%%FIGURE%%%%%%%%%%%%%%%%%%%%%%%%

The normalized LF mode dielectric strength $ \Delta\varepsilon_\mathrm{LF} 
/(c_\mathrm{HA} \cdot L_\mathrm{LF}^2)$ is shown in Fig.\ \ref{Fig7}: panel 
(a) displays its dependence on HA concentration in pure water solutions, and 
panel (b) dependence on 
the ionic strength of added salt ions for a representative HA concentration 
$c_\mathrm{HA}=0.03$~mg/mL. Again we assume 
that the dielectric strength was defined by the square of the length scale 
revealed by a given relaxation, and also by the concentration of the relaxation 
entities, \ie{}, fraction of counterions $f_\mathrm{LF}$ participating in the LF 
relaxation (see Eq.\ \ref{increment2}). In this manner the normalized dielectric strength presented in the 
panel (a) shows that the apparent fraction of counterions participating in the 
LF relaxation strongly decreases with an increase in HA concentration according 
to a power law $f_\mathrm{LF} \propto  \Delta\varepsilon_\mathrm{LF} 
/(c_\mathrm{HA} \cdot L_\mathrm{LF}^2) \propto c_\mathrm{HA}^{-0.37\pm0.04}$. 
This reflects the decrease in the dielectric strength 
$\Delta\varepsilon_\mathrm{LF}$ of the mode itself, since the product 
$(c_\mathrm{HA} \cdot L_\mathrm{LF}^2)$ is a constant due to the fact that 
$L_\mathrm{LF}$ is a renormalized Debye screening length which scales as 
$c_\mathrm{HA}^{-0.5}$  (see Eq.\ \ref{r_Bfc}). This result indicates that the concentration-independent Manning-based definition for the number of oscillating counterions is not valid in this case. Presumably, as the screening due to other polyions and counterions increases with 
concentration it might reduce the effective number of counterions 
participating in the LF relaxation along the renormalized Debye length $r_\mathrm{B}$. On the other hand, data displayed in panel (b) suggest that $f_\mathrm{LF}$ becomes larger in the case of added salt solutions compared to the pure water case, once the concentration of added salt ions becomes larger than $c_{\textrm{HA}}$ and relaxation happens along the electrostatic persistence length. A similar effect was observed for DNA solutions and ascribed to the intrinsic counterion atmospheres squeezed closer to the chains \cite{Tomic07}.

\section{Discussion}

First, we comment the issue of respective roles in screening of HA counterions
and ions from the added salt in the light of previously obtained results on long
DNA \cite{Tomic06,Tomic07}. For both systems the influence of added salt is
important as long as the added salt concentration $2I_\mathrm{s}$ is
sufficiently larger than the concentration of intrinsic counterions
$c_\mathrm{in}$. However, whereas for DNA we observed that these two regimes are
delimited at about $ {2I_\mathrm{s}}/{ c_\mathrm{in}}\approx 0.4$, for Na-Ha
this occurs at higher added salt levels ${2I_\mathrm{s}}/{ c_\mathrm{in}}\approx
1$. Since in the case of HA all intrinsic counterions are free, while there are
only 25\% free in the case of DNA, this result might indicate that the effective
counterion screening is due primarily to the free counterions and not to
condensed ones. Such a result is not surprising since the proper description of
strongly charged polyelectrolytes such as DNA, with large charge densities that
produce nonlinearities, is standardly tackled by taking into account the Manning
condensation while retaining the linearized Debye-H\"{u}ckel (DH) theory for the
remaining salt. Conversely, in the case of weakly charged polyelectrolytes, such
as HA, the DH approximation can be used without fundamental modifications.

Second, our understanding of the LF mode provides additional details in the
scenario for the HF mode. We remind that in the case of the long DNA the dGPD
correlation length gives way to the Debye screening length as a new relevant
length scale for HF mode. This occurs exactly when sufficient salt is added so
that the corresponding Debye screening length becomes comparable to and
eventually smaller than the dGPD correlation length \cite{Tomic07}. On the other
hand, in the case of HA the screening length is revealed as the pertinent length
scale for the LF mode. Only at very high added salt concentrations, the
screening length and the correlation length observed in the HF mode, will
theoretically become comparable.  For example, at ${ c_\mathrm{HA}}=0.03$~mg/ml
the correlation length $L_\mathrm{HF} \approx 25$~nm (see Fig.\ \ref{Fig4})
could become comparable to $L_\mathrm{LF}$ at added salt concentration
${I_\mathrm{s}}\approx 30$~mM [see Fig.\ \ref{Fig6}(b)]. Thus we expect the
correlation length to remain relevant for HF mode at the added salt
concentrations at least one order of magnitude higher than practically
measurable with our technique.

In what follows we address possible reasons for why the LF relaxation mode
observed for the HA polyelectrolyte yields $L_\mathrm{LF}$ scaling laws which
differ from the ones observed in the long DNA case \cite{Tomic06,Tomic07}. The
first issue concerns the salt-free regime in which the average size of the
Gaussian chain which scales as $c^{-0.25}$, previously observed for long DNA, is
replaced in the case of HA by the renormalized Debye screening length which
scales as $r_\mathrm{B} = C \cdot (B / b c_\mathrm{HA})^{0.5}$. We suggest that
the absence of Manning condensed counterions due to weaker electrostatic
interactions, characterized by a charge density parameter $\eta < 1$, might be
at the  origin of this observation. We base this suggestion on the experimental
data on long DNA, reported before, that have indicated that for the LF
relaxation in the salt-free regime it is mostly the condensed counterions that
oscillate along or in close proximity to an individual chain in the
polyelectrolyte solution, except at large enough salt concentrations where at
least some of the free counterions seem to join in in playing this role.
Contrary, in the case of HA only free counterions, Manning condensed being
absent, participate in the LF relaxation and their oscillation in the volume
around the chain naturally brings in the renormalized Debye screening length,
which has a stronger concentration dependence and therefore appears as a more
pertinent length scale than the average size of the chain.

The second issue is why for HA with finite added salt the characteristic length
scale of counterion dynamics, that for DNA allowed an association with the OSF
electrostatic persistence length, is replaced by the electrostatic persistence
length which scales linearly with Debye length $\kappa^{-1}$, \ie{}, as
$L_\mathrm{p} \propto I_\mathrm{s}^{-0.5}$. We consider it plausible that a much
higher flexibility of HA, as compared to DNA, might be responsible for this
change, so that the OSF approach in which the chain is modeled as a rigid rod
would be invalidated. This has been already stressed and discussed at length by
Ullner \cite{Ullner-book,Ullner03} and others \cite{Barrat93,Dobrynin-elperlen}.
Here we note that the controversy concerning the conformational properties of
flexible polyelectrolytes in the presence of added salt has been discussed over
the last fifty years without reaching a firm consensus \cite{Ullner-book}. In
particular, the electrostatic persistence length has been reported to depend
both
quadratically and linearly on the Debye screening length. The former is a result
of the analytical approaches based on the OSF-like perturbational calculation in
which the polyelectrolyte is modeled as a rigid rod, whereas the latter comes
out from the variational calculation for a Flory-type flexible chain. A weak
point in both approaches is the assumption that there is a single behavior that
describes the chain on all length scales. Recently, Ullner has argued
\cite{Ullner03}
that a consistent picture might emerge taking into account simulations with a
careful analysis of analytical approaches. According to his arguments, three
regimes can be distinguished with respect to the scaling of the persistence
length with the Debye screening length. In the first two regimes where
$\kappa^{-1}$ is larger or comparable to the size of the chain, $R$, the
chain thus being rather stiff, it behaves as a rod-like chain, so
that the OSF approach is valid and $L_\mathrm{p} \propto I_\mathrm{s}^{-1}$. In
the third regime where $\kappa^{-1}$ is small or the chain is weakly charged
\cite{Dobrynin-elperlen},
the chain is flexible enough to allow distant parts to get close, which gives
rise to increased long-range correlations so that the excluded volume effects
start playing an important role which affects the average size of the molecule
in a way that corresponds to a linear or even a sublinear dependence of the
persistence length on the ionic strength $I_\mathrm{s}$. There are various
expressions for the scaling exponent of the electrostatic persistence length on
the ionic strength in this regime, with the simplest one yielding behavior close
to
$I_\mathrm{s}^{-0.5}$ \cite{Ullner-book,Ullner03,Barrat93,Dobrynin-elperlen}.

Obviously these excluded volume effects are much less pronounced in the case of
DNA, which is intrinsically a stiff molecule. We find it difficult to apply the
criterion $\kappa^{-1}$ versus the chain size $R$ in the case of experimentally
studied system, since $\kappa^{-1}$ is always much smaller than $R$. Rather we
suggest that the comparison of the structural persistence length $L_0$ and the
correlation length $\xi$ should be more meaningful in this regard. Indeed, the
comparison of $L_0$ (50~nm for DNA and 9~nm for HA) and the correlation length
$\xi$ in the same concentration range (10--50~nm for DNA and 4--45~nm for HA)
indicates that while for DNA $L_0 \geq \xi$, for HA exactly the opposite
relationship holds, $L_0 \leq \xi$, indicating a much higher HA flexibility if
compared to DNA. This stronger flexibility of HA leads in its turn to a weaker
dependence of the electrostatic persistence length on added salt concentration
because of the more pronounced excluded volume effects. Indeed, by reanalyzing
viscosity data Tricot has demonstrated the validity of
$L_\mathrm{p} \propto I_\mathrm{s}^{-0.5}$ for different flexible
polyelectrolytes with structural persistence length between 2 and 20~nm
\cite{Tricot}.
In addition, he has shown that $L_\mathrm{p}$ does not diverge at vanishing
added salt concentration, rather it reaches the maximum value that may be much
smaller than the polymer extended length $L$. Additional confirmation to
corroborate these results comes from transient electric birifrigence
measurements performed on flexible Na-PSS for a range of low added salt
concentrations similar to ones used in our dielectric spectroscopy
measurements \cite{Degiorgio91}.

On the other hand, recent small angle neutron scattering (SANS) experiments on
HA for added salt concentrations larger than 1~mM, reveal the OSF behavior for
the electrostatic persistence length \cite{Buhler04,Bonnet08}. However, a closer
inspection of Fig.\ 10 in Ref.\ \cite{Buhler04} and of Fig.\ 3 in Ref.\ \cite{Bonnet08}
reveals that the apparent OSF behavior of data critically depends on the 0.1~M
(0.2~M) single data point which is not obtained from experiment, but is rather
calculated from the OSF formula. This result apparently contradicts our results
obtained by dielectric spectroscopy measurements, but only for added salt
concentrations smaller than 1~mM. We can conceive two possible reasons for this
discrepancy. The first cause might be associated with different added salt
regimes. Another cause might lie in the fact that the SANS experiments were done
on much shorter HA chains than our DS experiments, and the analysis of the SANS
data was based on theory in which excluded volume effects were neglected.
In future, more efforts are needed  in order to elucidate the crossover behavior
between low and high added salt regimes as far as HA and other flexible
polyelectrolytes are concerned. This task however appears to be a rather
difficult one for the following reason. The low added salt regime, below a few
mM, can be probed only by very few experimental techniques, which cannot be used
for high added salt concentrations, among which dielectric spectroscopy appears
as the most reliable one. The opposite holds for the high salt regime above 5~mM
or so: a number of techniques, while they cannot be used for low added salt, can
cover this regime ranging from viscosity measurements to SANS experiments. It
thus seems impossible to find a single technique for continuously following the
conformational change of the polyion over a broad added salt concentration
range.

All these results make it quite clear that the standard criterion to
discriminate flexible against semiflexible polyelectrolytes, which compares
the structural persistence length with  the monomer size, cannot be always
strictly applied. It also appears difficult to apply the criterion $\kappa^{-1}$
\vs{}\ $R$ as argued by Ullner {\it et al.}\ \cite{Ullner03}. Another
criterion which considers the structural persistence length in comparison with
the correlation length appears to be of key importance which determines, at
least for added salt concentrations below few mM, the outcome of two competing
electrostatic persistence lengths: the one scaling linearly with the Debye
length
versus the one scaling quadratically with the Debye length, \ie{}, the
Odijk-Skolnick-Fixman length. Keeping this in mind, regarding the salt
dependence of its persistence length the HA seems to be somewhere in between the
semiflexible DNA and the flexible Na-PSS.

\section{Conclusion}

In conclusion, our results demonstrate that in the case of low added salt there 
are basically two fundamental length scales that determine the dielectric 
response of semidilute solutions of semiflexible polyelectrolyte HA: de 
Gennes-Pfeuty-Dobrynin semidilute solution correlation length and de 
Gennes-Dobrynin electrostatic screening length. The first length scale is 
important for the high frequency response, while the second one is important for 
the low frequency response. Collective properties probed by the high frequency 
dielectric relaxation are thus well described by the de Gennes-Pfeuty-Dobrynin 
solution correlation length predicted and proven to be valid for both flexible 
and semiflexible chains. Single-chain properties probed by the low frequency 
relaxation are described by the electrostatic persistence length that scales 
linearly with the Debye length, contrary to the DNA case where the scaling is 
quadratic as rationalized by the OSF formula. This difference is most probably 
due to the fact that the HA chain is  much more flexible than the DNA chain, 
precluding the straightforward application of the OSF arguments. This also 
indicates that HA, as far as its flexibility is concerned, is half-way between 
semiflexible and completely flexible polyelectrolytes. In the salt-free regime 
the de Gennes-Dobrynin screening length is due to HA counterions only, which can 
thus be described by a renormalized Debye screening length that apparently 
prevails as the fundamental single-chain property probably due to the absence of 
the condensed counterions in the weakly charged polyelectrolyte. Finally, our 
results reveal that the standard theoretical approaches based on the 
concentration-independent Manning parameter $\eta$ face some limitations when 
applied to the dielectric response of  the counterion atmosphere around 
polyions. These theories describe well the HF dielectric relaxation along the 
correlation length, with its concentration-independent fraction of oscillating 
counterions $f=1/\eta$. On the other hand, they fail to offer a proper description of the 
LF relaxation characterized by the renormalized Debye screening length. For this 
relaxation, the effective number of counterions decreases with increasing 
concentration, which might be ascribed to enhanced screening due to other 
polyions and counterions.

\begin{acknowledgments}
We would like to thank P.\ A.\ Pincus for a valuable and illuminating discussion.
This work was supported by the Croatian Ministry of Science, Education and
Sports under grant 035-0000000-2836 (Strongly correlated inorganic, organic and
biomaterials). R.P. would like to acknowledge the
financial support by the Slovenian Research Agency under contract
P1-0055 (Biophysics of Polymers, Membranes, Gels, Colloids and Cells) and
J1-0908(Active media nanoactuators with dispersion forces). This study was in
part supported by the Intramural Research Program of the NIH, Eunice Kennedy
Shriver National Institute of Child Health and Human Development.
\end{acknowledgments}


\begin{thebibliography}{99}

\bibitem{Daune03}
M.\ Daune,
{\em Molecular Biophysics} (Oxford University Press, New York, 2003).

\bibitem{Bloomfield00}
V.\ A.\ Bloomfield, D.\ M.\ Crothers, and I.\ Tinocco, Jr.,
{\em Nucleic Acids} (University Science Books, Sausalito, 2000).

\bibitem{Elias}
H.\ G.\ Elias,
{\em Macromolecules} (Plenum Press, New York, 1984).

\bibitem{Bordi}
F.\ Bordi, C.\ Cametti, and R.\ H.\ Colby,
J.\ Phys.: Condens.\ Matter {\bf 16}, R1423 (2004).

\bibitem{Ito}
K.\ Ito, A.\ Yagi, N.\ Ookubo, and R.\ Hayakawa,
Macromolecules {\bf 23}, 857 (1990).

\bibitem{Tomic06}
S.\ Tomi\'{c}, T.\ Vuleti\'{c}, S.\ Dolanski Babi\'{c}, S.\ Kr\v{c}a, D.\ Ivankovi\'{c}, L.\ Gripari\'{c}, and R.\ Podgornik,
Phys.\ Rev.\ Lett.\ {\bf 97}, 098303 (2006).

\bibitem{Tomic07}
S.\ Tomi\'{c}, S.\ Dolanski Babi\'{c}, T.\ Vuleti\'{c}, S.\ Kr\v{c}a, D.\ Ivankovi\'{c}, L.\ Gripari\'{c}, and R.\ Podgornik,
Phys.\ Rev.\ E\ {\bf 75}, 021905 (2007).

\bibitem{Tomic08}
S.\ Tomi\'{c}, S.\ Dolanski Babi\'{c}, T.\ Ivek, T.\ Vuleti\'{c}, S.\ Kr\v{c}a,
F.\ Livolant, and R.\ Podgornik,
Europhys.\ Lett.\ {\bf 81}, 68003 (2008).

\bibitem{Nandi}
N.\ Nandi, K.\ Bhattacharyya, B.\ Bagchi,
Chem.\ Rev.\  (Washington, D.C.) {\bf 100}, 2013 (2000).

\bibitem{Buhler04}
E.\ Buhler and F.\ Boue,
Macromolecules {\bf 37}, 1600 (2004).

\bibitem{deGennes76}
P.\ G.\ de Gennes, P.\ Pincus, R.\ M.\ Velasco, and F.\ Brochard,
J.\ Phys.\ (Paris) {\bf 37}, 1461 (1976).

\bibitem{note1}
$c^\ast$ is given by the concentration where there is only one polymer molecule
in the volume of a polymer globule $c^\ast=\textrm{molecule mass}/V_\mathrm{c}$,
where molecule mass is $N \cdot m_\mathrm{m}$, and
$V_\mathrm{c} \approx L_\mathrm{c}^3 = N^3 \cdot b^3$;
$m_\mathrm{m}$ is a mass of a monomer $\approx 401$~g/mol, $N=4000$,
$L_\mathrm{c}=N \cdot b$; $b=1$~nm; $b$ is the monomer size.

\bibitem{Schwan}
H.\ P.\ Schwann, G.\ Schwarz, J.\ Maczuk, and H.\ Pauly, J.\ Phys.\ Chem.\ {\bf
66}, 2626 (1962).

\bibitem{Essex}
C.\ G.\ Essex, G.\ P.\ South, R.\ J.\ Sheppard, and E.\ H.\ Grant, J.\ Phys.\ E:
Sci.\ Instrum.\ {\bf 8}, 385 (1975).

\bibitem{Agilent}
H.\ Haruta, {\em The Impedance Measurement Handbook, 2nd edition} (Agilent
Technologies, USA/Japan, 2000).

\bibitem{Holling}
A.\ D.\ Hollingsworth, and D.\ A.\ Saville, J.\ Colloid Interface Sci.\ {\bf
257}, 65 (2003).


\bibitem{Roldan}
R.\ Roldan-Toro, and J.\ D.\ Sollier, J.\ Colloid Interface Sci.\ {\bf 274}, 76
(2004).

\bibitem{Schwan68}
H.\ P.\ Schwan, and C.\ D.\ Ferris, Rev.\ Sci.\ Instrum.\ {\bf 39}, 481 (1968).

\bibitem{Davey}
C.\ L.\ Davey, G.\ H.\ Markx, and D. B. Kell, Eur.\ Biophys.\ J.\ {\bf 18}, 255
(1990).

\bibitem{Saif91}
B.\ Saif, R.\ K.\ Mohr, C.\ J.\ Montrose and T.\ A.\ Litovitz, Biopolymers {\bf 31}, 
1171 (1991).

\bibitem{Grosse}
C.\ Grosse, M.\ Tirado, W.\ Pieper, and R.\ Pottel, J.\ Coll.\ Interface Sci.\ {\bf 205}, 26 (1998).

\bibitem{Osawa71}
F.~Oosawa, {\em Polyelectrolytes}, (Marcel Dekker, New York, 1971).

\bibitem{OBrian86}
R.\ W.\ O'Brian, J.\ Coll.\ Interface Sci.\ {\bf 113}, 81 (1986).

\bibitem{Wong}
T.\ E.\ Angelini, R.\ Golestanian, R.\ H.\ Coridan, J.\ C.\ Butler, A.\ Beraud, M.\ Krisch, H.\ Sinn, K.\ S.\ Schweizer, and G.\ C.\ L.\ Wong,
Proc.\ Natl.\ Acad.\ Sci.\ USA {\bf 103}, 7962 (2006).

\bibitem{Bordi02}
F.\ Bordi, C.\ Cametti, and T.\ Gili,
Phys.\ Rev.\ E\ {\bf 66}, 021803 (2002).

\bibitem{Odijk79}
T.\ Odijk,
Macromolecules {\bf 12}, 688 (1979).

\bibitem{Pfeuty78}
P.\ Pfeuty,
J.\ Phys.\ (Paris) {\bf 39}, C2-149 (1978).

\bibitem{Dobrynin}
A.\ V.\ Dobrynin and M.\ Rubinstein,
Prog.\ Polym.\ Sci.\ {\bf 30}, 1049 (2005);
A.\ V.\ Dobrynin, R.\ H.\ Colby, and M.\ Rubinstein,
Macromolecules {\bf 28}, 1859 (1995).


\bibitem{note2}
The lower accuracy of our data for added salt concentrations above 0.2~mM are
due to strong influence of the electrode polarization effects.


\bibitem{note3}
Data obtained on the long and short 146~bp DNA fragments give 50~nm value for
the structural persistence length and 50~nm value for the contour length,
respectively.

\bibitem{Odijk77}
T.\ Odijk,
J.\ Polym.\ Sci.: Polym.\ Phys.\ {\bf 15}, 477 (1977).

\bibitem{Skolnick77}
J.\ Skolnick and M.\ Fixman,
Macromolecules {\bf 10}, 944 (1977).

\bibitem{Ullner-book}
M.\ Ullner,
{\em Polyelectrolytes. Physicochemical Aspects and Biological Significance}, in
{\em DNA Interactions with Polymers and Surfactants}, edited by R.\ Dias and B.\
Lindman (Wiley-Interscience, 2008).

\bibitem{Ullner03}
M.\ Ullner,
J.\ Chem.\ Phys.\ B {\bf 107}, 107 (2003).

\bibitem{Barrat93}
J.\ L.\ Barrat and J.\ F.\ Joanny,
Europhys.\ Lett.\ {\bf 24}, 333 (1993).

\bibitem{Dobrynin-elperlen}
A.\ V.\ Dobrynin,
Macromolecules {\bf 38} 9304 (2005).

\bibitem{Tricot}
M.\ Tricot,
Macromolecules {\bf 17}, 1698 (1984).

\bibitem{Degiorgio91}
V.\ Degiorgio, F.\ Mantegazza, and R.\ Piazza,
Europhys.\ Lett.\ {\bf 15}, 75 (1991).

\bibitem{Bonnet08}
F.\ Bonnet, R.\ Schweins, and F.\ Boue,
Europhys.\ Lett.\ {\bf 83}, 48002 (2008).

\end{thebibliography}
\end{document}